# Study of one-step and two-step neutron transfer in the reaction $^6$Li + $^9$Be

A.K. Azhibekov[1,2], S.M. Lukyanov[1], Yu.E. Penionzhkevich[1,3], B.A. Urazbekov[4], M.A. Naumenko[1], V.V. Samarin[1,5], T. Issatayev[1,4,6], V.A. Maslov[1], K. Mendibayev[*,1,6], D. Aznabayev[1], T.K. Zholdybayev[6], A. Temirzhanov[6,7]

[1]Joint Institute for Nuclear Research, Dubna, Russia
[2]Korkyt Ata Kyzylorda University, Kyzylorda, Kazakhstan
[3]National Research Nuclear University MEPhI, Moscow, Russia
[4]L.N. Gumilyov Eurasian National University, Astana, Kazakhstan
[5]Dubna State University, Dubna, Russia
[6]Institute of Nuclear Physics, Almaty, Kazakhstan
[7]Satbayev University, Almaty, Kazakhstan
*E-mail address: kayrat1988@bk.ru

**Abstract:** The paper presents the results of experiments on measuring the cross sections for elastic scattering and nucleon transfer channels in the $^6$Li+$^9$Be reaction at an incident energy of 68 MeV: $^9$Be($^6$Li,$^6$Li)$^9$Be, $^9$Be($^6$Li,$^7$Li)$^8$Be, $^9$Be($^6$Li,$^7$Li)$^8$Be$_{2+}$, $^9$Be($^6$Li,$^8$Li)$^7$Be, $^9$Be($^6$Li,$^7$Be)$^8$Li. The aim is to reveal the manifestation of the cluster structure of $^9$Be. Theoretical analysis of the contributions of one-step and two-step neutron transfer mechanisms is performed using the distorted wave Born approximation method with the Fresco code. Good agreement between the calculations and the experimental data was obtained for the channels of elastic scattering $^9$Be($^6$Li,$^6$Li)$^9$Be, neutron $^9$Be($^6$Li,$^7$Li)$^8$Be and proton transfer $^9$Be($^6$Li,$^7$Be)$^8$Li, as well as the transfer of two neutrons $^9$Be($^6$Li,$^8$Li)$^7$Be. It is shown that the dineutron cluster transfer mechanism makes a dominant contribution to the $^9$Be($^6$Li,$^8$Li)$^7$Be reaction channel at forward angles.

## 1. Introduction

The study of one-step and two-step transfer of nucleons and clusters in nuclear reactions provides a possibility of answering the question of the existence of multi-neutron systems which is one of the important problems of nuclear physics. The problem of the existence of light neutron clusters (dineutron, tetraneutron, etc.) is more than 60 years old, but it is still of interest for both theoretical and experimental research [1-5]. A recent paper [6] reported that the observation of a resonance structure near the threshold for the formation of a four-neutron system corresponds to a quasi-bound tetraneutron cluster that manifests itself in the reaction $^8$He+$p$→$^4$He+$p$+$^4n$ and lives for a very short time.

Concerning the problem of studying light neutron clusters, a dineutron is of great interest. The dineutron can be formed near the surface of neutron-rich nuclei [7]. The first attempt to observe the unstable dineutron, i.e., a system of two neutrons in the singlet state, was made by V.K. Voitovetskii et al. in the reaction $^2$H($n,p$)$^2n$ at $E_n$ = 14 MeV from the spectrum of the protons [8]. In [9], the authors studied the decay of the $^{16}$Be nucleus and obtained the results indicating the dineutron nature of the decay with a small angle of emission between two neutrons. The measured two-neutron separation energy for $^{16}$Be was 1.35 MeV, which was consistent with calculations in the shell model. However, in the calculations, the authors did not take into account the interaction between the emitted neutrons which can also explain the observed correlations of the emission angles of the two neutrons [10].

Two-neutron transfer reactions are a unique tool for studying the interaction between neutrons and confirming the existence of the dineutron clusters that manifest themselves during the interaction of two nuclei [11-15]. The difficulty in interpreting experimental data is that such reactions can proceed as both one-step and two-step neutron transfer processes which cannot be separated experimentally [16]. In [17], the authors showed that the product nuclei are formed as a result of one-step transfer of two neutrons and that the contribution of this process is especially significant when low-lying excited states are populated in the formed nuclei. Thus, to describe experimental data, it is extremely important to take into account the probabilities of both one-step and two-step neutron transfer [18].

In the elastic scattering of $^6$He nuclei on $^4$He [19-22] and $^4$He on $^6$Li [23], an increase in the cross section at backward angles was observed. The authors interpreted this increase as the existence of the dineutron cluster in the $^6$He nucleus and the deuteron cluster in the $^6$Li nucleus. However, the optical model of elastic scattering could not describe this effect, while the calculation of the corresponding transfer cross sections for dineutron and deuteron clusters within the framework of the distorted wave Born approximation (DWBA) method fully explained this behavior as the contribution of the channel of the transfer of a two-nucleon cluster.

Another interesting experimental result was obtained in [24] for the $^6$He+$^{65}$Cu reaction at the beam energy of 22.6 MeV: the cross section for the two-neutron transfer was found to be larger than that for one-neutron transfer. From this, the authors concluded that "the dineutron configuration of $^6$He plays a dominant role in the reaction mechanism."

Concerning the $^9$Be nucleus, it was revealed that the dineutron cluster manifests itself in the reaction channel $^9$Be($^3$He,$^7$Be)$^5$He [25, 26]. The $^9$Be($^3$He,$^6$He)$^6$Be reaction channel observed at forward angles corresponds to the transfer of three neutrons. The calculations done in [25] within the framework of the coupled reaction channels (CRC) method showed that the two-step transfer mechanisms ($n$-$^2n$ and $^2n$-$n$) make a significant contribution to the cross section, which is also indirect evidence of the transfer of the dineutron cluster.

This work is a part of our systematic studies of nucleon and cluster transfer in reactions with various projectiles on $^9$Be: $d$+$^9$Be [27], $^3$He+$^9$Be [25, 26]. Here we study reaction channels with the weakly bound projectile nucleus $^6$Li. The aim is to reveal the manifestation of the cluster structure of $^9$Be in the studied reaction channels. In particular, we focus on the reaction channel $^9$Be($^6$Li,$^8$Li)$^7$Be in order to estimate whether a one-step or two-step transfer is the most probable mechanism of the transfer of two neutrons.

Section 2 provides a detailed description of the experiment. Section 3 presents experimental cross sections and their comparative analysis. Sections 4 and 5 are devoted to the theoretical method employed to analyze the experimental data. Sections 6 and 7 present the results of the theoretical analysis of the experimental data.

## 2. Experiment

The experiment was performed at the Flerov Laboratory of Nuclear Reactions, Joint Institute for Nuclear Research, Dubna. An intense $^6$Li beam with an energy of 68 MeV was accelerated by the U-400 cyclotron and transported to the reaction chamber (Fig. 1) of the high-resolution magnetic analyzer MAVR [28].

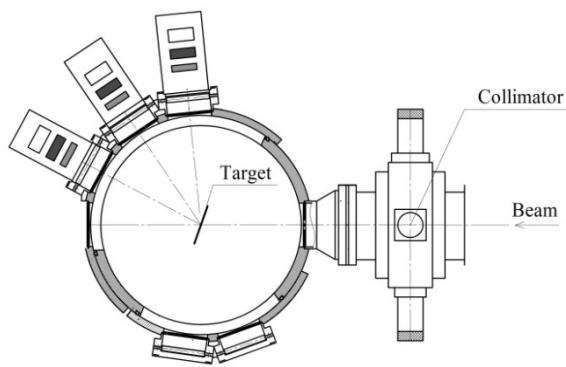

**Fig. 1.** The reaction chamber with the $^9$Be foil target and three three-layer semiconductor telescopes ($\Delta E_1$, $\Delta E_2$, $E_R$).

The beam profile was formed by the magnetic optics of the U-400 cyclotron supplemented by a system of diaphragms. The beam size was controlled by the profilometer installed in front of the reaction chamber; on a target, it was 5 × 5 mm$^2$ at an intensity of 30 nA. The total number of particles passing through the target was determined by a Faraday cup and also monitored by elastic scattering.

The beam was focused onto the self-supporting 5-μm thick $^9$Be foil. The target purity was better than 99%; a possible admixture of carbon and oxygen isotopes in the target material was not observed in the measured energy spectra.

Particle identification was done by measuring energy losses and residual energy in detectors (Δ$E$-$E$ method). For this purpose, three three-layer semiconductor telescopes were used, the first two thin detectors of which measured specific energy losses Δ$E_1$, Δ$E_2$ (Fig. 1). Their thickness was 50 μm and 300 μm, respectively. The third detector $E_R$ was 3.2-mm thick and measured the residual energy of the reaction products after they passed through the first two detectors. The configuration of such telescopes made it possible to reliably identify reaction products from helium to boron isotopes in a wide energy range. Examples of identification matrices obtained by one of the telescopes used in the experiments are shown in Fig. 2. It can be seen that the reaction products were unambiguously identified.

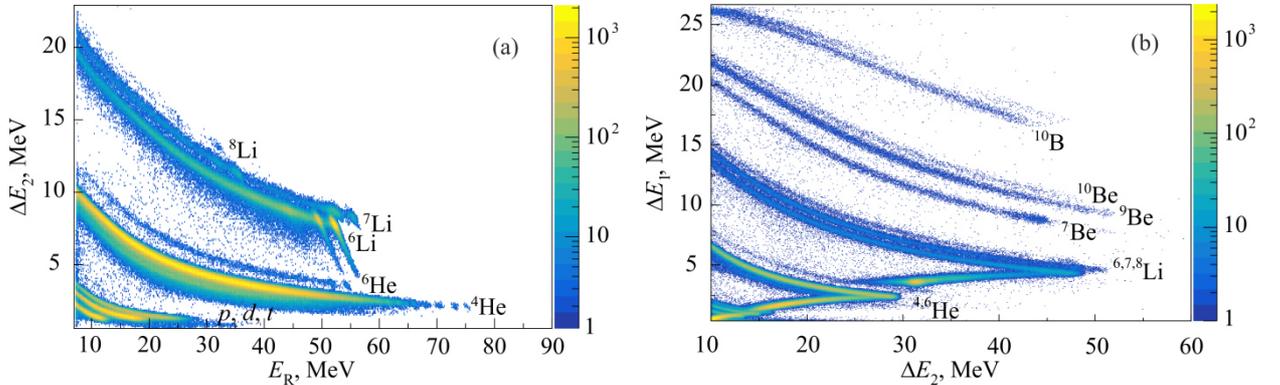

**Fig. 2.** Examples of identification matrices obtained by one of the telescopes: (a) Δ$E_2$-Δ$E_R$ at an angle of 16° and (b) Δ$E_1$-Δ$E_2$ at 12°.

To measure the energy spectra and angular distributions of the nuclei emitted in the reaction, we used an inclusive method. The energy resolution of the detecting system was determined by the energy resolution of the $^6$Li beam and errors in measuring the energy losses of particles in the target material. In the case of registration of particles with $Z$ = 1–3, the energy resolution was ≈ 500 keV; for particles with $Z$ = 4–5, it was ≈ 1 MeV.

The excitation energy spectra corresponding to the energy of the states of the $^9$Be, $^8$Be, $^7$Be, and $^8$Li nuclei are shown in Figs. 3 and 4. The populated ground and first excited states are indicated. It should be noted that the width of each peak of a state in the spectra was determined by several factors: natural width, instrument resolution of the spectrometer, and energy spread. Events corresponding to multibody exit channels make insignificant contributions to these peaks. In the resulting energy spectra, we can see the excited states of the complementary products corresponding to the two-body exit channels. The complementary products are

– $^9$Be in the reaction channel $^9$Be($^6$Li,$^6$Li)$^9$Be (in the case of detection of $^6$Li [Fig. 3(a)]);
– $^8$Be in the reaction channel $^9$Be($^6$Li,$^7$Li)$^8$Be (in the case of detection of $^7$Li [Fig. 3(b)]);
– $^7$Be in the reaction channel $^9$Be($^6$Li,$^8$Li)$^7$Be (in the case of detection of $^8$Li) [Fig. 4(a)];
– $^8$Li in the reaction channel $^9$Be($^6$Li,$^7$Be)$^8$Li (in the case of detection of $^7$Be [Fig. 4(b)]).

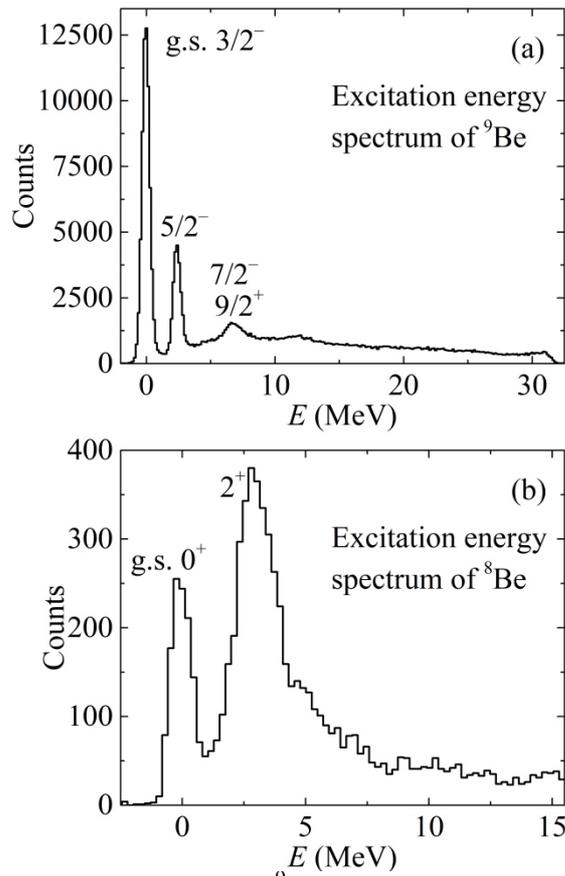

**Fig. 3.** Excitation energy spectra for (a) $^9$Be in the case of detection of $^6$Li at an angle of 16° and (b) $^8$Be in the case of detection of $^7$Li at 14°.

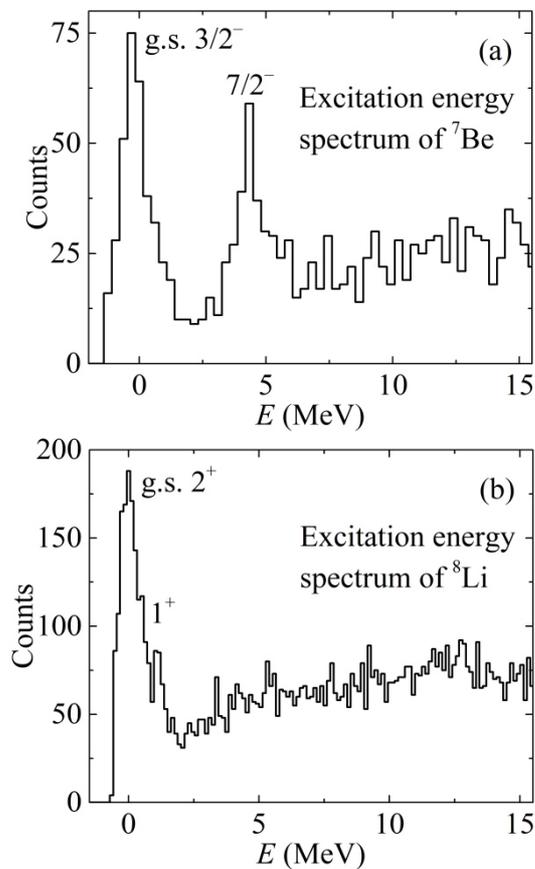

**Fig. 4.** Excitation energy spectra for (a) $^7$Be in the case of detection of $^8$Li at an angle of 12° and (b) $^8$Li in the case of detection of $^7$Be at 16°.

The narrow peak at 2.43 MeV in Fig. 3(a) is the first excited rotational level $5/2^-$ of the $^9$Be nucleus, the next wide peak corresponds to overlapping of two excited levels ($7/2^-$, 6.38 MeV) and ($9/2^+$, 6.76 MeV) of $^9$Be; the narrow peak at 3.03 MeV in Fig. 3(b) is the first excited rotational level $2^+$ of the $^8$Be nucleus. For the $^9$Be and $^8$Be nuclei, the first rotational levels $5/2^-$ and $2^+$ are populated with large probabilities. The single-particle excited levels of the $^9$Be nucleus are not observed because of the small neutron separation threshold (Table 1). For the $^7$Be nucleus [Fig. 4(a)], the first low-lying single-particle excited level ($1/2^-$, 0.43 MeV) is not separated from the ground state peak, the second single-particle excited level ($7/2^-$, 4.57 MeV) is observed. Other single-particle excited levels of the $^7$Be nucleus are not observed because they are above the proton separation threshold (Table 1) and are populated with low probability. For the $^8$Li nucleus [Fig. 4(b)], the first low-lying single-particle excited level ($1^+$, 0.98 MeV) is situated near the ground state peak. Other single-particle excited levels of the $^8$Li nucleus are not observed because they are above the neutron separation threshold (Table 1) and are populated with low probability.

**Table 1.** Particle separation energies for the $^8$Li and $^{7,8,9}$Be nuclei.

| Nucleus | Alpha particle α (MeV) | Proton $p$ (MeV) | Neutron $n$ (MeV) |
|---------|------------------------|------------------|-------------------|
| $^8$Li  | 6.1                    | 12.4             | 2.03              |
| $^7$Be  | 1.59                   | 5.6              | 10.7              |
| $^8$Be  | –0.092                 | 17.3             | 18.9              |
| $^9$Be  | 2.46                   | 16.9             | 1.66              |

### 3. Angular distributions of reaction products

Differential cross sections for each angle were obtained taking into account solid angles of the telescopes, the thickness of the target, and the number of particles incident on the target. The experimental setup made it possible to measure the energy spectra of the reaction products in the range of angles 10°–83° in the laboratory system; the error in measuring the angle was ±1°. Angles greater than 83° could not be measured because of the design features of the scattering chamber. The measured angular distributions of the products of the reaction $^6$Li+$^9$Be at an energy of 68 MeV are presented in Figs. 5, 6. The relative error in cross-section measurements is not larger than 20%. This error is predominantly due to the following factors: statistical errors in counting events, errors in target thickness determination, inaccuracies in solid angle values, and errors in beam intensity measurements.

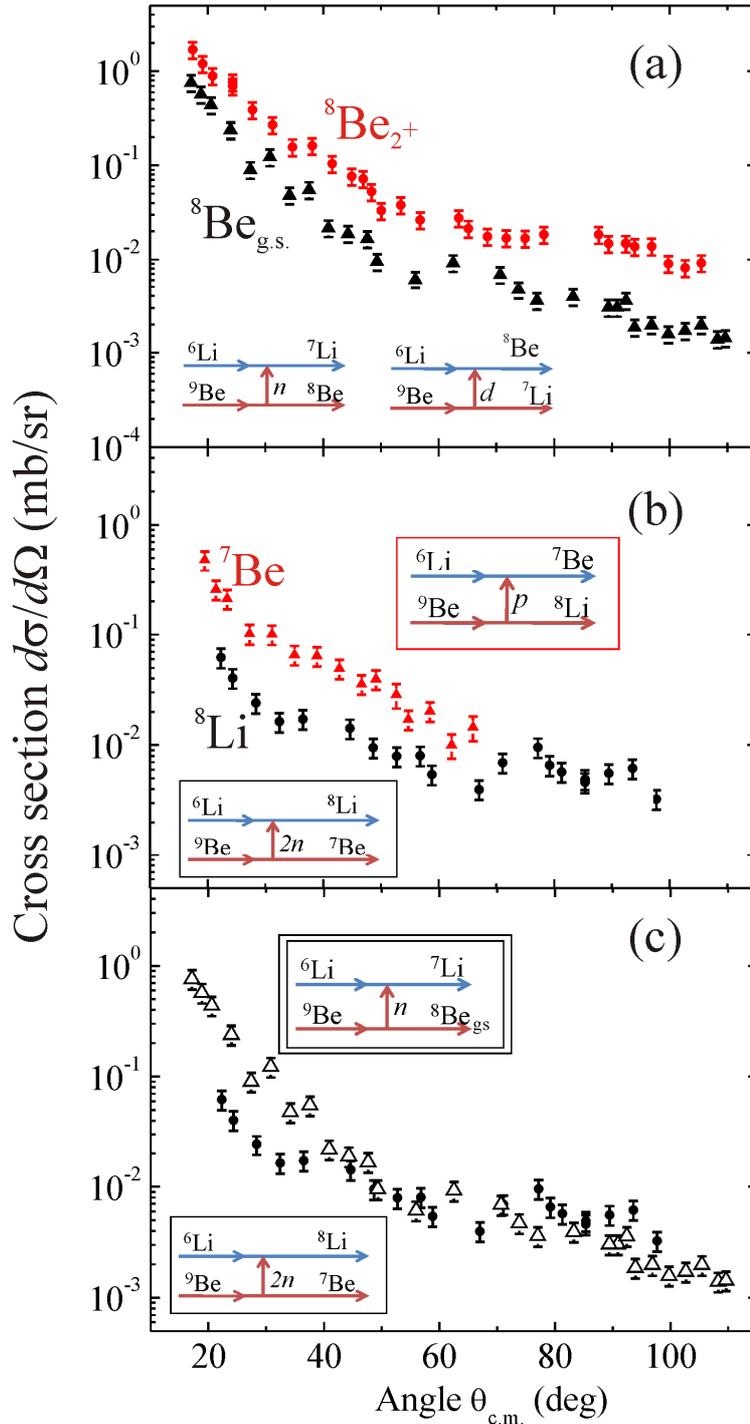

**Fig. 5.** Experimental angular distributions for the products of the $^6$Li+$^9$Be reaction at 68 MeV: (a) $^7$Li in exit channels $^7$Li+$^8$Be$_{g.s.}$ (triangles) and $^7$Li+$^8$Be$_{2+}$ (circles); (b) $^8$Li (circles) and $^7$Be (triangles) in exit channels $^8$Li$_{g.s.}$+$^7$Be$_{g.s.,1/2-}$ and $^7$Be$_{g.s.}$+$^8$Li$_{g.s.,1+}$, respectively; (c) $^7$Li in exit channel $^7$Li+$^8$Be$_{g.s.}$ (empty triangles) and $^8$Li (circles) in exit channel $^8$Li$_{g.s.}$+$^7$Be$_{g.s.,1/2-}$. Transfer mechanisms are shown in insets.

The relatively high cross sections for the transfer of one neutron $n$ from the weakly bound target nucleus $^9$Be [Fig. 5(a)] can be explained by the manifestation of its cluster structure ($\alpha + n + \alpha$) [27]. The differential cross sections for the transfer of two neutrons in the reaction channel $^9$Be($^6$Li,$^8$Li)$^7$Be have comparable values with those for the reaction channels of the transfer of one neutron $^9$Be($^6$Li,$^7$Li)$^8$Be$_{g.s.}$ and $^9$Be($^6$Li,$^7$Li)$^8$Be$_{2+}$. As can be seen from Fig. 5(c), the ratio $\sigma_{1n}/\sigma_{2n}$ for the reaction channels $^9$Be($^6$Li,$^7$Li)$^8$Be$_{g.s.}$ and $^9$Be($^6$Li,$^8$Li)$^7$Be at forward angles is approximately equal to 10 and smoothly decreases with increasing angle to ~ 1.

## 4. Theoretical analysis of elastic scattering

Experimental differential cross sections for elastic scattering of $^6$Li on the $^9$Be nucleus are presented in Fig. 6. The experimental data were analyzed within the optical model using the Fresco code [29, 30]. The optical potential used in the calculations was represented as

$$U(r) = -V_V f(r; R_V, a_V) - iV_W f(r; R_W, a_W) + V_C(r), \qquad (1)$$

with the Woods-Saxon form-factors for both the real and imaginary parts

$$f(r; R_{V,W}, a_{V,W}) = \left\{1 + \exp\left[(r - R_{V,W})/a_{V,W}\right]\right\}^{-1}, \qquad (2)$$

where

$$R_{V,W} = r_{V,W}\left(A_p^{1/3} + A_t^{1/3}\right), \qquad (3)$$

$V_V$ and $V_W$ are the depth parameters for the real and imaginary parts of the optical potential, respectively; $r_{V,W}$ and $a_{V,W}$ are geometric parameters; $A_p$ and $A_t$ are the mass numbers of the projectile and target nuclei, respectively; $V_C(r)$ is the Coulomb potential of a uniformly charged sphere with the radius

$$R_C = r_C\left(A_p^{1/3} + A_t^{1/3}\right). \qquad (4)$$

In our calculations, we used the parameter $r_C = 0.717$ fm.

The theoretical elastic scattering cross section was fitted to the measured experimental data within the optical model using the SFresco code [30]. As a starting point for seeking the optical potential in our calculations, we took the parameters for the elastic scattering of $^6$Li+$^9$Be at an energy of 50 MeV [31]. All six parameters, the depths $V_{V,W}$ and the geometric parameters $r_{V,W}$, $a_{V,W}$, were varied. It can be seen that we achieved an excellent fit ($\chi^2/N=1.418$) of the experimental data (Fig. 6). The parameters of potential (1) are listed in Table 2.

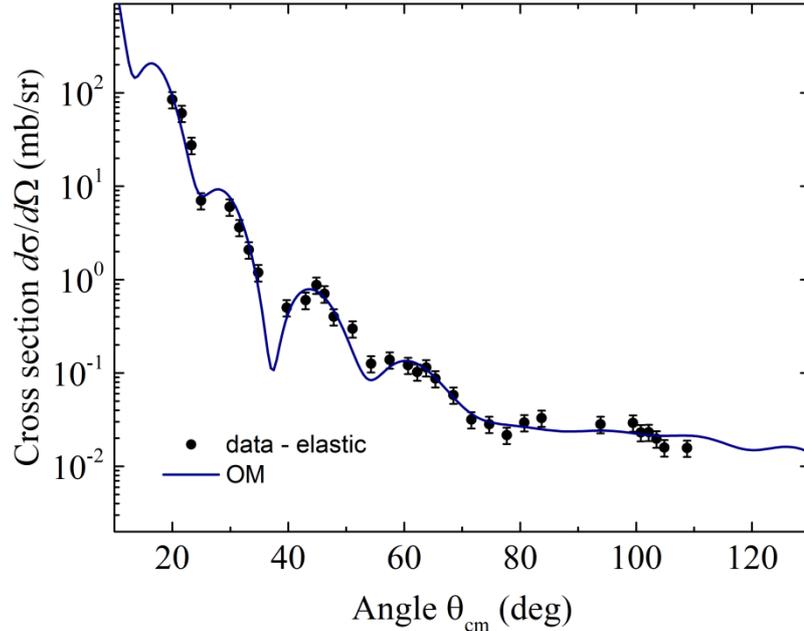

**Fig. 6.** Angular distributions for the elastic scattering channel $^9$Be($^6$Li,$^6$Li)$^9$Be: experimental data (circles) and results of calculations (curves).

Table 2. Parameters of the optical potential (1) for elastic scattering.

| $V_V$ (MeV) | $r_V$ (fm) | $a_V$ (fm) | $V_W$ (MeV) | $r_W$ (fm) | $a_W$ (fm) |
|---|---|---|---|---|---|
| 152.20 | 0.698 | 0.624 | 12.36 | 1.388 | 0.930 |

## 5. Theoretical analysis of transfer channels

Theoretical analysis of the cross sections for the transfer channels was performed within the DWBA method [32, 33] using the Fresco code [29, 30]. We calculated the one-step transfer using the prior formalism of the DWBA amplitude. For the two-step transfer of two neutrons, we used the second-order DWBA; a prior-post combination was chosen to avoid non-orthogonality terms [30, 32, 33]. According the DWBA formalism, the main ingredients required for calculations are the internal wave functions $(\phi_A, \phi_b)$, $(\phi_a, \phi_B)$ for the nuclei in the transfer reaction $A+b \rightarrow a+B$ ($A=a+x$, $B=b+x$). The wave function for the nucleus $B$ with the total spin $J$ and spin projection $M$ can be written as [30, 32, 33]

$$\phi_B^{JM}(\xi, r) = \sum_{Ilj} A_{lsj}^{IJ} \left[ \phi_b^I(\xi) \otimes \varphi_{lsj}(r) \right]_{JM}, \qquad (5)$$

where the coefficients $A_{lsj}^{IJ}$ are the so called coefficients of fractional parentage (CFP) or spectroscopic amplitudes, and their square moduli $S_{lsj}^{IJ} = \left| A_{lsj}^{IJ} \right|^2$ are the spectroscopic factors [30, 32, 33]. Spectroscopic factors can be considered as a probability of finding the nucleon or cluster $x$ in a single-particle state with the quantum numbers $l, s, j$ bound to the core $a$ or $b$ with the spin $I$ [32]. Below, we denote the spectroscopic amplitude as $A_x$, where $x$ is the nucleon or cluster with all its quantum numbers. All spectroscopic amplitudes $A_x$ used in our calculations (Table 3) were taken from the shell model calculations published in [13, 25, 34-37].

Table 3. Spectroscopic amplitudes $A_x$ for the nucleon or cluster $x$ in the systems $A = a + x$ or $B = b + x$ [13, 25, 34-37].

| $A$ or $B$ | $a$ or $b$ | $x$ | $nl_j$ | $A_x$ |
|---|---|---|---|---|
| $^7$Li | $^6$Li | $n$ | $1p_{3/2}$ | –0.735 |
| $^7$Li$_{0.477}$ | $^6$Li | $n$ | $1p_{3/2}$ | 0.329 |
| $^8$Li | $^6$Li | $^2n$ | $1d_2$ | –0.667 |
| $^8$Li | $^7$Li | $n$ | $1p_{3/2}$ | –0.478 |
| $^7$Be | $^6$Li | $p$ | $1p_{3/2}$ | –0.735 |
| $^7$Be$_{0.429}$ | $^6$Li | $p$ | $1p_{3/2}$ | –1.740 |
| $^8$Be | $^7$Be | $n$ | $1p_{3/2}$ | –1.234 |
| $^8$Be$_{3.03}$ | $^7$Be | $n$ | $1p_{3/2}$ | 0.771 |
| $^8$Be$_{3.03}$ | $^7$Be$_{0.429}$ | $n$ | $1p_{3/2}$ | –0.655 |
| $^9$Be | $^8$Li | $p$ | $1p_{1/2}$ | –0.375 |
| $^9$Be | $^8$Be | $n$ | $1p_{3/2}$ | 0.866 |
| $^9$Be | $^7$Be | $^2n$ | $2s_0$ | 0.247 |

The wave functions of the bound states of the nucleons and dineutron clusters $x$ in the target and projectile nuclei were obtained using the Woods-Saxon potential. The potential depths were adjusted to reproduce the experimental binding energies of the nucleons and clusters [38], while the parameters $a$ and $r_0$ were fixed: $a = 0.65$ fm and $r_0 = 1.25A^{1/3}$ fm [13, 25, 39].

We adjusted only the potential parameters for the exit channels [40], while keeping the parameters for the entrance and intermediate channels, as well as the spectroscopic amplitudes.

## 6. Reaction channel $^9$Be($^6$Li,$^7$Li)$^8$Be

Experimental differential cross sections for the $^9$Be($^6$Li,$^7$Li)$^8$Be$_{gs}$ and $^9$Be($^6$Li,$^7$Li)$^8$Be$_{3.03}$ channels in comparison the DWBA calculations are shown in Fig. 7. For these reaction channels, we detected the $^7$Li nucleus, and the $^8$Be product was considered complementary to the detected one.

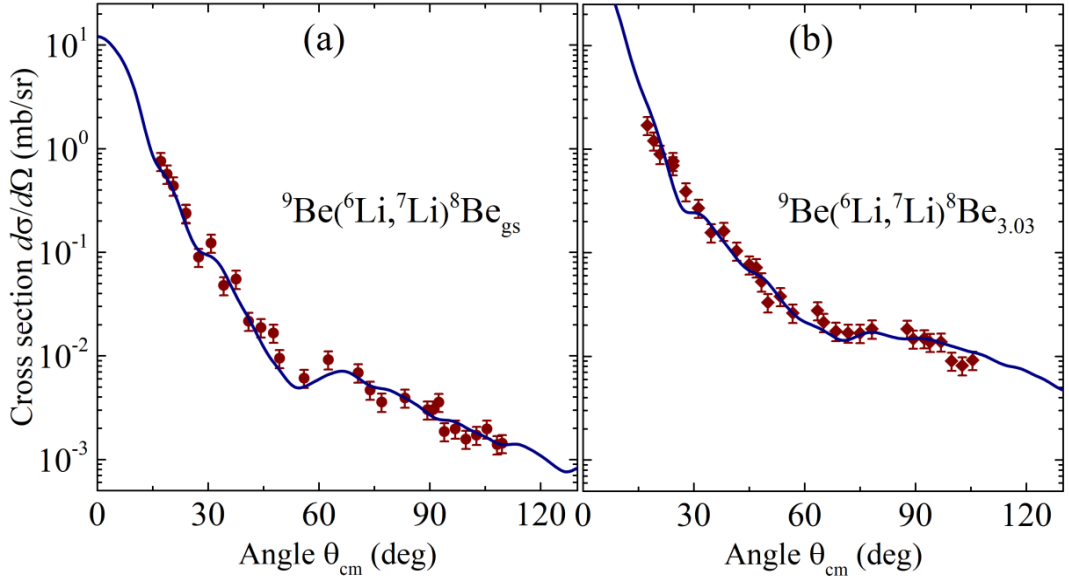

**Fig. 7.** Experimental angular distributions for the neutron transfer channels (a) $^9$Be($^6$Li,$^7$Li)$^8$Be$_{gs}$ and (b) $^9$Be($^6$Li,$^7$Li)$^8$Be$_{3.03}$ (symbols) in comparison with the results of calculations (curves).

The potential parameters describing the elastic scattering of $^6$Li+$^9$Be (Table 2) were used for the entrance channel. For the $^7$Li+$^8$Be$_{g.s.}$ exit channel, we also used a potential in the Woods-Saxon form with the parameters obtained by fitting the calculation results to the experimental data on the angular distributions. The parameters recommended in [41] were used as initial parameters in the fitting procedure.

The parameters of the real part of the potential for the $^7$Li+$^8$Be$_{g.s.}$ exit channel were the same as for the $^7$Li+$^8$Be$_{3.03}$ exit channel. However, due to the fact that the values of the cross sections for the $^9$Be($^6$Li,$^7$Li)$^8$Be$_{3.03}$ reaction channel are higher, we reduced the depth parameter $V_W$ of the imaginary part for the $^7$Li+$^8$Be$_{3.03}$ exit channel. To better reproduce the shape of the experimental angular distributions, we fitted the radius parameter $r_W$ of the imaginary part. As a result, an increased value of the radius parameter $r_W$ was obtained for the $^7$Li+$^8$Be$_{3.03}$ exit channel. It can be seen that our DWBA calculations reproduce the experimental data well (Fig. 7). The obtained parameters of the optical potential are given in Table 4.

**Table 4.** Parameters of the optical potential (1) obtained for the specified exit channels.

| Reaction channel | $V_V$ (MeV) | $r_V$ (fm) | $a_V$ (fm) | $V_W$ (MeV) | $r_W$ (fm) | $a_W$ (fm) | $r_C$ (fm) |
|---|---|---|---|---|---|---|---|
| $^7$Li+$^8$Be | 152.20 | 0.669 | 0.853 | 30.50 | 1.008 | 0.809 | 0.677 |
| $^7$Li+$^8$Be$_{3.03}$ | 152.20 | 0.669 | 0.853 | 12.36 | 1.388 | 0.809 | 0.677 |

## 7. Reaction channels $^9$Be($^6$Li,$^7$Be)$^8$Li and $^9$Be($^6$Li,$^8$Li)$^7$Be

Similar to the $^9$Be($^6$Li,$^7$Li)$^8$Be reaction channel, calculations were carried out for the $^9$Be($^6$Li,$^7$Be)$^8$Li reaction channel where the proton transfer occurs from the target nucleus to the projectile nucleus, leading to the exit channel $^7$Be+$^8$Li. The spectroscopic amplitudes of the transferred proton used for the configurations $^6$Li+$p$ and $^8$Li+$p$ are given in Table 3. The wave functions of the protons in the nuclei $^7$Be = $^6$Li+$p$ and $^9$Be = $^8$Li+$p$ were calculated by varying the depth of the Woods–Saxon potential to reproduce the binding energy exactly in the same way as it was done in the previous section. It is worth mentioning that the binding energy of the proton $p$ in the $^9$Be nucleus is 16.9 MeV, which is comparable to the binding energy of two neutrons $^2n$ in the $^9$Be nucleus, 20.6 MeV.

The parameters of the Woods-Saxon potential for the $^7$Be+$^8$Li$_{gs}$ and $^7$Be+$^8$Li$_{0.98}$ exit channels (Table 5) were obtained by fitting the calculation results to the experimental angular distribution at fixed potential parameters for the entrance channel ($^6$Li + $^9$Be, Table 2). The parameters from [42] were used as initial parameters in the fitting process.

**Table 5.** Parameters of potential (1) obtained for the specified exit channels.

| Reaction channel | $V_V$ (MeV) | $r_V$ (fm) | $a_V$ (fm) | $V_W$ (MeV) | $r_W$ (fm) | $a_W$ (fm) | $r_C$ (fm) |
|---|---|---|---|---|---|---|---|
| $^7$Be+$^8$Li$_{gs}$ | 125.50 | 0.657 | 0.853 | 12.25 | 0.888 | 0.809 | 0.664 |
| $^7$Be+$^8$Li$_{0.98}$ | 115.04 | 0.657 | 0.853 | 12.36 | 0.888 | 0.809 | 0.664 |

The measured energy spectrum of $^7$Be for the $^9$Be($^6$Li,$^7$Be)$^8$Li reaction channel is shown in Fig. 4(b). Because of the relatively low energy resolution, the low-lying excited state of $^8$Li ($1^+$, 0.98 MeV) is practically not resolved from the ground state. The results of the DWBA calculations for the transfer of a proton in the reaction channels $^9$Be($^6$Li,$^7$Be)$^8$Li$_{gs,0.98}$ are shown in Fig. 8. They are fairly close to the experimental angular distribution of the $^7$Be nucleus. The solid blue curve in Fig. 8 represents an incoherent sum of the cross sections for the ground state and the first excited state 0.98 of $^8$Li. The significant contribution of the reaction channel $^9$Be($^6$Li,$^7$Be)$^8$Li$_{0.98}$ in the incoherent sum of the cross sections is consistent with the measured energy spectrum of $^7$Be shown in Fig. 4(b).

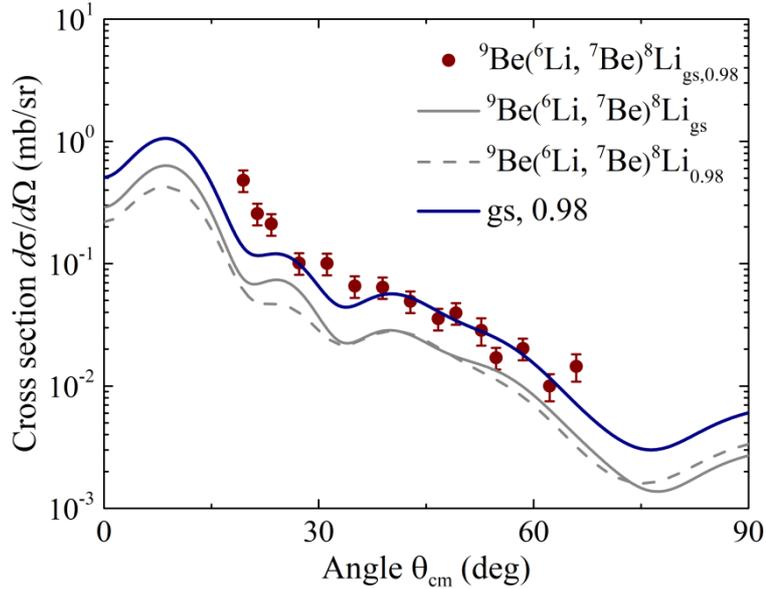

**Fig. 8.** Experimental angular distribution of the reaction channels $^9$Be($^6$Li,$^7$Be)$^8$Li$_{gs,0.98}$ (circles) in comparison with the results of calculations (curves).

The values of the experimental differential cross sections at forward angles (three points at $\theta_{cm}$=19.5-23.4°) could not be described by the theoretical curve; this may be due to the presence of a contribution from the inelastic excitation of the $^6$Li projectile [43]. Nevertheless, the shape of the theoretical curve in this angular range is close to the experimental data.

A similar discrepancy between DWBA calculations and experimental data was observed in [43, 44]. This can be overcome by adjusting the values of the spectroscopic amplitudes or reducing the imaginary part of the exit channel potential, but in this case, we will lose the overall agreement of the DWBA calculations with the experimental points. This fact indicates the need for more complex calculations taking into account inelastic excitations of the nuclei in the studied channels; this is the subject of a separate future theoretical work.

The proton transfer, inverted relative to the scattering angle, serves as an alternative to the two-neutron transfer mechanisms in the $^9\text{Be}(^6\text{Li},^8\text{Li})^7\text{Be}$ reaction channel [45]. Therefore, we included it in the calculations shown in Fig. 9. The one-step mechanism corresponds to the transfer of a di-neutron cluster $^2n$ or a proton $p$, while the two-step mechanism – to the two-step transfer of two neutrons $n$-$n$.

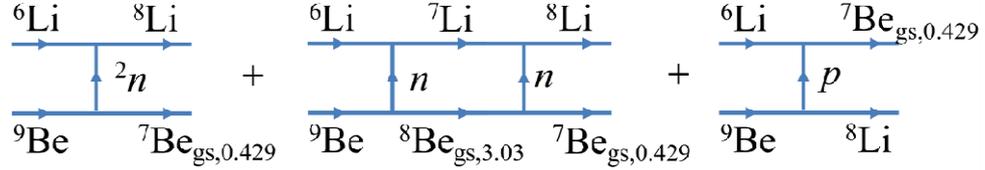

**Fig. 9.** Reaction mechanisms taken into account in calculations.

We used the same potential parameters from Table 5 for the $^7\text{Be}+^8\text{Li}_{gs}$ and $^8\text{Li}_{gs}+^7\text{Be}$ exit channels. The potential parameters obtained in Sections 5 and 6 were used for the entrance channel ($^6\text{Li}+^9\text{Be}$, Table 2) and for the intermediate channels ($^7\text{Li}+^8\text{Be}_{gs}$ and $^7\text{Li}+^8\text{Be}_{3.03}$ in calculations for the $n$-$n$ transfer, Table 4). The potential parameters for the exit channel $^8\text{Li}+^7\text{Be}_{0.429}$ were obtained by fitting the cross section to reproduce the experimental data. As a starting point for seeking the potential, we used the parameters for the exit channel $^7\text{Li}+^8\text{Be}$ (Table 4). The resulting parameters were as follows: $V_V$= 155 MeV, $r_V$ = 0.669 fm, $a_V$=0.853 fm, $V_W$ = 19.25 MeV, $r_W$ = 1.388 fm, $a_W$ = 0.780 fm.

The spectroscopic amplitudes $A_x$ for the nucleons and dineutron clusters included in the calculations are listed in Table 3. The higher value of $A_x$ = 0.667 for the configuration $^8\text{Li}=^6\text{Li}+2n$ compared to $A_x$ = 0.478 for the configuration $^8\text{Li}=^7\text{Li}+n$ may favor dineutron cluster transfer mechanism (i.e., one-step transfer of two neutrons) compared to the two-step mechanism of neutron transfer.

The differential cross section for the $^9\text{Be}(^6\text{Li},^8\text{Li})^7\text{Be}$ reaction channel has the form of a coherent sum of two amplitudes

$$\frac{d\sigma}{d\Omega}(\theta_{cm}) = \left| f_I(\theta_{cm}) + f_{II}(\theta_{cm}) \right|^2, \quad (6)$$

where $f_I(\theta_{cm})$ and $f_{II}(\theta_{cm})$ are the amplitudes of one-step and two-step transfer mechanisms, respectively [29, 30]:

$$f_I(\theta_{cm}) = f_{2n}(\theta_{cm}) + f_p(\pi - \theta_{cm}), \; f_{II}(\theta_{cm}) = f_{n-n}(\theta_{cm}), \quad (7)$$

$f_{2n}(\theta_{cm})$ and $f_p(\pi - \theta_{cm})$ are the amplitudes of the one-step transfer of two neutrons and proton, respectively; $f_{n-n}(\theta_{cm})$ is the amplitude of the two-step transfer of two neutrons. The experimental angular distribution of the $^8\text{Li}$ nucleus for the reaction channel $^9\text{Be}(^6\text{Li},^8\text{Li})^7\text{Be}$ in comparison with the calculation results is shown in Fig. 10(a). In Fig. 10(b), the DWBA calculations for each mechanism taken into account in the coherent sum (6) are presented separately:

– for one-step transfer:
- $^6\text{Li}+^9\text{Be} \rightarrow {}^8\text{Li}_{gs}+^7\text{Be}_{gs,0.429}$ ($^2n$–1 curve) – dineutron transfer included in the Sum-1 curve;
- $^6\text{Li}+^9\text{Be} \rightarrow {}^8\text{Li}_{gs}+^7\text{Be}_{gs}$ ($^2n$–2 curve) – dineutron transfer included in the Sum-2 curve;
- $^6\text{Li}+^9\text{Be} \rightarrow {}^7\text{Be}_{gs,0.429}+^8\text{Li}_{gs}$ ($p$–1 curve) – proton transfer included in the Sum-1 curve;
- $^6\text{Li}+^9\text{Be} \rightarrow {}^7\text{Be}_{gs}+^8\text{Li}_{gs}$ ($p$–2 curve) – proton transfer included in the Sum-2 curve;

– for two-step *n-n* transfer:
- $^6$Li+$^9$Be → $^7$Li$_{gs}$+$^8$Be$_{gs}$ → $^7$Be$_{gs,0.429}$ +$^8$Li$_{gs}$ (*nn*–1 curve) included in the Sum-1 curve;
- $^6$Li+$^9$Be → $^7$Li$_{gs}$+$^8$Be$_{gs}$ → $^7$Be$_{gs}$+$^8$Li$_{gs}$ (*nn*–2 curve) included in the Sum-2 curve;
- $^6$Li+$^9$Be → $^7$Li$_{gs}$+$^8$Be$_{gs,3.03}$ → $^7$Be$_{gs,0.429}$ +$^8$Li$_{gs}$ (*nn*–3 curve) included in the Sum-3 curve.

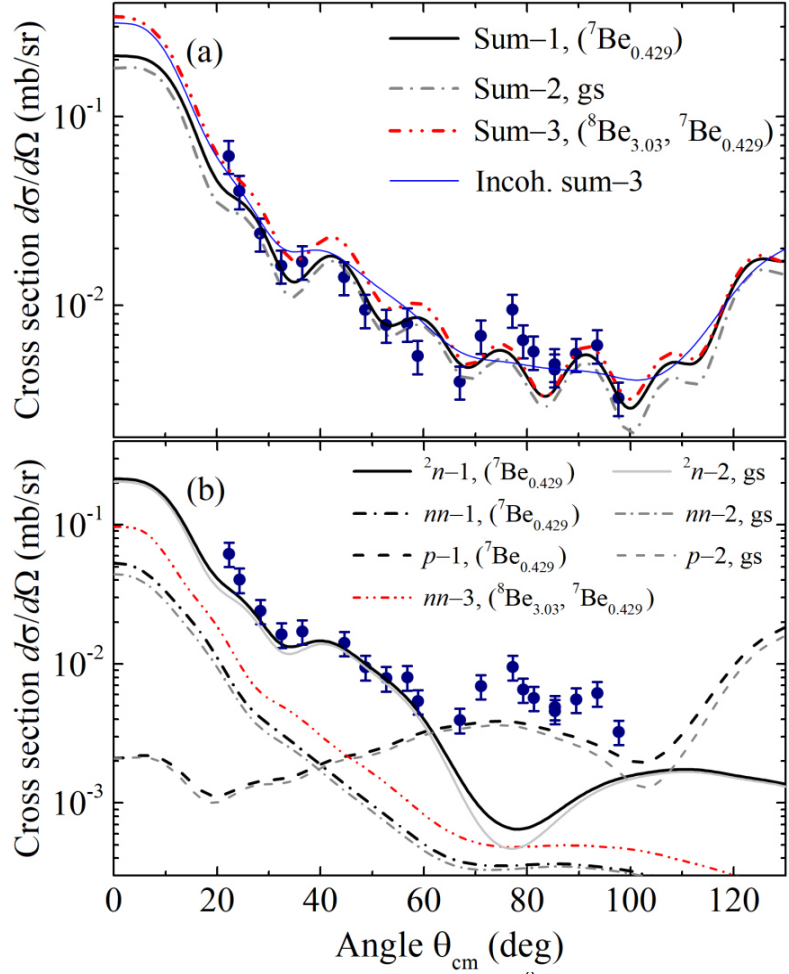

**Fig. 10.** Experimental angular distribution of the $^8$Li nucleus for the nucleon transfer channels $^9$Be($^6$Li,$^8$Li)$^7$Be$_{gs,0.429}$ (circles) in comparison with the results of calculations. The curves are the results of calculations including the excited states given in parentheses. (a) Sums of reaction mechanisms. (b) Contributions of nucleon and cluster transfer mechanisms taken into account in the calculations.

The angular distributions of the $^9$Be($^6$Li,$^8$Li)$^7$Be$_{gs,0.429}$ reaction channels in Fig. 10(a) have oscillations. These oscillations indicate the interference of the transfer mechanisms presented in Fig. 9. The dineutron transfer makes a relatively large contribution at forward angles, while in the range of angles 60–130°, the proton transfer dominates [Fig. 10(b)]. The contribution of two-step neutron transfer is negligible in the entire range of angles, which is consistent with the results of [13, 18, 46, 47].

Similar oscillations due to the interference of two-neutron and α-transfer mechanisms in the reaction $^{14}$C($^{16}$O, $^{18}$O)$^{12}$C were obtained in [45]. The incoherent sum yields a smoother angular distribution [45, 48], as shown by the thin solid curve in Fig. 10(a). It can be assumed that the characteristic features of two-neutron transfer reactions are the domination of the one-step transfer of two neutrons and the interference of the transfer mechanisms [18, 46, 47, 49, 50].

Taking into account the $^8$Be$_{3.03}$ state in the intermediate channel of two-step neutron transfer provides better description of the experimental points in the region of $\theta_{cm}$=20°-30°, but at the same time the calculated cross sections in the region of $\theta_{cm}$=45-70° are slightly overestimated [Fig. 10(a)]. The excitation of the $^8Be_{3.03}$ state in the intermediate channel does not have a strong effect on the calculation results for the $^9Be(^6Li,^8Li)^7Be$ reaction channel. The same results were obtained in [13] for $^9Be(^7Be,^9Be)^7Be$.

The contribution of the $^9Be(^6Li,^8Li)^7Be_{0.429}$ channel is insignificant, which is consistent with the excitation energy spectrum of the $^7Be$ nucleus [Fig. 4(a)]. However, taking into account the excitation of $^7Be_{0.429}$ in the exit channel improves the theoretical description of the cross sections [Fig. 10(a)]. We also note that if we exclude all excitations from the calculations, the experimental differential cross sections will be underestimated, for example, in the regions of angles $\theta_{cm}=0-25°$ and $\theta_{cm}=70-85°$, which indicates the importance of taking into account excitations of nuclei in the reaction.

Typically, the structure of the $^8Li$ nucleus is considered in the two-body $^7Li+n$ or three-body $\alpha+n+t$ theoretical models [51, 52]. The $^9Be$ nucleus is usually represented as a system of two α-clusters and a neutron located with a high probability between them $\alpha+n+\alpha$ [53-55]. However, the large contribution of the dineutron transfer to the cross sections [Fig. 10b] suggests that the $^8Li$ and $^9Be$ nuclei can manifest configurations corresponding to the two-body structures $^6Li+^2n$ and $^7Be+^2n$, respectively. The contribution of the dineutron transfer mechanism in the range of angles 0–60° is approximately 5 times higher than the contribution of the mechanism of two-step transfer of two neutrons [46].

**Conclusions**

The energy and angular distributions for the $^9Be(^6Li,^6Li)^9Be$, $^9Be(^6Li,^7Li)^8Be$, $^9Be(^6Li,^8Li)^7Be$, and $^9Be(^6Li,^7Be)^8Li$ channels in the $^6Li+^9Be$ reaction at an energy of 68 MeV were measured. The energy distributions of the detected nuclei reproduce the scheme of population of the ground and low-lying excited states of complementary nuclei, which confirms the two-body nature of the reaction exit channels considered in this work.

From the analysis of the experimental data on elastic scattering of $^6Li+^9Be$, the parameters of the Woods-Saxon optical potential were determined. Taking into account elastic scattering as well as one- and two-step transfer reaction mechanisms allowed us to obtain good agreement with the experimental data on the $^9Be(^6Li,^8Li)^7Be$ reaction channel. It was shown that the dineutron transfer $^2n$ makes a larger contribution to the cross sections of the $^9Be(^6Li,^8Li)^7Be$ reaction channel at forward angles compared to proton transfer and two-step transfer of two neutrons. The proton transfer makes a contribution comparable to the dineutron transfer in the range of angles 60-130°. The contribution of two-step transfer of two neutrons is negligible in the entire range of angles. The oscillations in the angular distribution for the $^9Be(^6Li,^8Li)^7Be$ reaction channel indicate the interference of the transfer mechanisms. The large contribution of the dineutron transfer to the cross sections suggests that the $^8Li$ and $^9Be$ nuclei can manifest configurations corresponding to the two-body structures $^6Li+^2n$ and $^7Be+^2n$, respectively.

**Acknowledgments**

This research was funded by the Science Committee of the Ministry of Science and Higher Education of the Republic of Kazakhstan (Grant No. AP19677087) and by the Russian Science Foundation (Grant No. 24-22-00117).